# Activating the fluorescence of a Ni(II) complex by energy transfer

Tzu-Chao Hung[a,b], Yokari Godinez-Loyola[c,d], Manuel Steinbrecher[a], Brian Kiraly[a], Alexander A. Khajetoorians[a], Nikos L. Doltsinis[e], Cristian A. Strassert[c,d,f], and Daniel Wegner[a*]

[a] Institute for Molecules and Materials, Radboud University, 6500 GL Nijmegen, The Netherlands

[b] Institute for Experimental and Applied Physics, University of Regensburg, 93040 Regensburg, Germany

[c] Institut für Anorganische und Analytische Chemie, University of Münster, 48149 Münster, Germany

[d] Center for Nanotechnology (CeNTech), University of Münster, 48149 Münster, Germany

[e] Institut für Festkörpertheorie and Center for Multiscale Theory and Computation, University of Münster, 48149 Münster, Germany

[f] Cells in Motion Interfaculty Centre (CiMIC) and Center for Soft Nanoscience (SoN), University of Münster, 48149 Münster, Germany

[*] Email: d.wegner@science.ru.nl

**Abstract**

Luminescence of open-shell 3*d* metal complexes is often quenched due to ultrafast intersystem crossing (ISC) and cooling into a dark metal-centered excited state. We demonstrate successful activation of fluorescence from individual nickel phthalocyanine (NiPc) molecules in the junction of a scanning tunneling microscope (STM) by resonant energy transfer from other metal phthalocyanines at low temperature. By combining STM, scanning tunneling spectroscopy, STM-induced luminescence, and photoluminescence experiments as well as time-dependent density functional theory, we provide evidence that there is an activation barrier for the ISC, which in most experimental conditions is overcome. We show that this is also the case in an electroluminescent tunnel junction where individual NiPc molecules adsorbed on an ultrathin NaCl decoupling film on a Ag(111) substrate are probed. However, when placing an MPc (M = Zn, Pd, Pt) molecule close to NiPc by means of STM atomic manipulation, resonant energy transfer can excite NiPc without overcoming the ISC activation barrier, leading to Q-band fluorescence. This work demonstrates that the thermally activated population of dark metal-centered states can be avoided by a designed local environment at low temperatures paired with a directed molecular excitation into vibrationally cold electronic states. Thus, we can envisage the use of luminophores based on more abundant transition metal complexes that do not rely on Pt or Ir by restricting vibration-induced ISC.

**Introduction**

Molecular luminescence (e.g. fluorescence and phosphorescence) is a ubiquitous phenomenon, providing fundamental insights into light-matter interaction and the electronic and dynamic properties of molecules upon excitation.[1] Porphyrins and phthalocyanines are a group of chromophores that have historically gained particular attention.[2-4] In general, the fluorescence from the lowest ligand-centered singlet $(\pi,\pi^*)$ excited state ($S_1$) competes with radiationless deactivation pathways, including intersystem crossing (ISC) and internal conversion (IC). While IC is impaired for highly rigid closed-shell complexes, ISC is promoted by heavy central atoms and becomes faster as spin-orbit coupling increases. However, the population of dark states that formally involve the occupation of antibonding metal-centered $d^*$ orbitals promotes radiationless deactivation pathways by conical intersection with the ground state. This is a general problem encountered for open-shell 3d transition metal complexes (and somewhat less for 4d elements), with the metallophthalocyanines FePc, CoPc and NiPc and some of its derivatives being studied particularly well.[2,5-12] In case of NiPc, the deactivation of luminescence was explained by the fact that excitation to the $S_1$ state is followed by ultrafast (< 1 ps) ISC to a vibrationally hot metal-centered $(d,d^*)$ state, eventually leading to nonradiative decay to the $S_0$ ground state within a lifetime of about 300 ps.[7,8,12] This efficient conversion of electronic excitation into heat can be used in photothermal therapy and photoacoustic imaging.[13,14] In optoelectronics, however, luminescent complexes generally rely on expensive and rare elements such as Pt or Ir, providing intrinsically high *d*-orbital splitting paired with high SOC to promote phosphorescence.[15] Clearly, a sustainable display technology should rely on less critical elements, which is why vast efforts are performed to chemically design complexes accommodating abundant 3d elements,[16,17] especially Cu(I)[18,19] and Ni(II).[20-22] Here, highly rigid luminophoric ligands with optimized ligand-field splittings are used, while avoiding the population of dissociative states by pushing the antibonding $d^*$ orbitals up in energy.

While the established strategy is to tweak the intramolecular structure chemically toward optoelectronic applicability,[23] an interesting alternative approach may be the physical design of intermolecular interactions, by controlling the local environment around the luminophore. In this context, recent studies combining scanning tunneling microscopy (STM) with light detection, referred to as STM-induced luminescence (STML) spectroscopy,[24] not only enabled to fundamentally understand single-molecule fluorescence with submolecular resolution,[25-32] but especially the influence of the local environment and neighboring chromophores could be investigated using STM-based atomic manipulation techniques. For example, the impact of adsorption at defects and step edges of an insulating surface on the fluorescence spectrum were studied,[33] exciton delocalization and superradiance were observed in J-aggregated ZnPc dimers and chains,[25,34,35] and resonant energy transfer (RET) in molecular donor-acceptor dimers and trimers was investigated with atomic-scale resolution.[36-38] The latter is particularly interesting, as it may offer the opportunity to activate luminescence in otherwise dark open-shell 3d-metal complexes by physical design of the local environment rather than intramolecular chemical design.

In this context, we chose to use NiPc for a proof-of-principle experiment, as the MPc family is well established in STML experiments, permitting stable thermal deposition onto clean substrates in an ultrahigh vacuum environment. We show that fluorescence of individual NiPc molecules can be activated at low temperatures through resonant energy transfer from neighboring MPc molecules (M = Zn, Pd, Pt). To demonstrate this, we combined STM, atomic manipulation, scanning tunneling spectroscopy (STS) and STML to identify individual NiPc and MPc molecules, to assemble them into NiPc-MPc dimers, and to perform spatially resolved fluorescence spectroscopy. Using three monolayers (ML) NaCl grown on Ag(111) as a substrate, the molecules were sequentially deposited on the surface and their orbital energies and structure characterized. We observed no luminescence from individual NiPc molecules, while Q-band emission was detected when NiPc was dimerized with another MPc and the latter was excited via the STM tunnel current. From spatially dependent STS and STML as well as a comparison of dimers with various intermolecular separations, we determined that RET is responsible for the excitation of NiPc. We explain the emergence of fluorescence by a deactivation of ISC, which is caused by a combination of freezing the relevant molecular vibrations and providing insufficient energy in the RET process to overcome the ISC activation barrier. Thus, the results suggest that a local environment at the cold interface providing precisely tuned energy funneling into the vibrationally "cold" electronic state responsible for the emission of light enables the rather rare emission from an otherwise "dark" 3d-metal complex.

**Results**

We first characterized the structural, optical and electronic properties of NiPc and all MPc monomers by means of STM, STML and STS at $T$ = 4.5 K. Fig. 1a shows STM images (inset) as well as STML spectra of individual NiPc, PtPc, PdPc and ZnPc molecules adsorbed on 3 ML NaCl/Ag(111). The STM topography images taken at a sample bias voltage $V_s$ = –2.5 V looked similar for all molecules, mostly reflecting the spatial distribution of the highest occupied molecular orbital (HOMO) (see also spatial maps in Fig. 1b). Note that the image of ZnPc reflects the superposition of two different adsorption orientations, as the molecule rapidly shuttles between them;[33] contrastingly, NiPc, PtPc, and PdPc did not show rapid shuttling, and the aromatic rings are oriented along the NaCl ⟨110⟩ crystallographic directions. However, one peculiar difference seen for NiPc was an increased noise level in the constant-current topography. This was not seen on the other MPc molecules, even when measured with the same tip and identical imaging parameters (see Fig. 3). As this apparent instability was reproduced using more than a dozen different microtips, this observation cannot be a tip artifact but is intrinsic to NiPc. We note that the noisy appearance was observed for all voltages, and a histogram analysis did not show discrete telegraph noise-like steps that would be indicative of switching events.

STML spectra reproduced the single-molecule fluorescence observed in previous studies of ZnPc[25,29,33,34], PdPc[36] and PtPc[38] adsorbed on ultrathin NaCl films, illustrating that the tip used is properly calibrated for STML (Fig. 1a). The most prominent spectral feature is the < 20 meV-sharp Q(0,0) resonance emitted upon relaxation of the given MPc from the $S_1$ excited singlet state to the $S_0$ ground state, with a maximum at $E(Q_{PtPc})$ = 1.94 eV, $E(Q_{PdPc})$ = 1.91 eV and $E(Q_{ZnPc})$ = 1.90 eV, for each of the respective molecules. In contrast, NiPc did not show any Q-band emission (top spectrum in Fig. 1a). To verify this, we also checked STML spectra at other sample bias voltages $V_s$, including at positive polarity (see Supplementary Information, Section 2), but we could not find any emission for NiPc in all studied parameter space.

The different character of NiPc compared to the other MPc molecules is also reflected in STS. As shown in Fig. 1b, d$I$/d$V$ spectra of the other MPc molecules looked similar, with one peak visible below the Fermi level ($E_F$) at $V_s < -2$ V and a second peak found above ($E_F$) at $V_s \approx 1$ V. These peaks are the positive (PIR) and negative ion resonances (NIR) when tunneling out of the HOMO or into the lowest unoccupied molecular orbital (LUMO), respectively. This was further confirmed by d$I$/d$V$ maps taken at the respective peak voltages, reflecting the spatial distributions of the HOMO and LUMO (see inset images in Fig. 1b). The electronic HOMO-LUMO gap (determined by identifying the onset energies[27]) varied from 2.85 eV (ZnPc) to 3.0 eV (PtPc) and to 3.05 eV (PdPc). NiPc also displayed two peaks in STS, but at very different voltages. The onset of the PIR was found at $V_s = -1.30$ V (maximum at $-1.6$ V), significantly closer to $E_F$ than for the other MPc molecules. The NIR onset was located at $V_s = 1.55$ V (maximum at 1.9 V). Hence, the electronic HOMO-LUMO gap of NiPc was identical to that of ZnPc, but the peaks were shifted higher in energy by about 1 eV. We will discuss the consequences of this for STML further below. We discuss a possible explanation for the relatively large shift of the NiPc d$I$/d$V$ spectrum in the Supporting Information (Section 2).

In order to confirm that the lack of fluorescence from NiPc is not related to the STML technique, we performed separate optical absorption and photoluminescence spectroscopy of monomer ensembles of the derivative NiPc-($t$Bu)$_4$ and MPc-($t$Bu)$_4$ in solution at room temperature (Fig. 2). We note that the *tert*-butyl groups were necessary to improve the solubility of the molecules, but as they are located at β-positions of the Pc macrocycle (see inset in Fig. 2a for the molecular structure), they should not significantly affect the optical properties.[39] From the absorption spectra (Fig. 2a), the Q-band of NiPc-(*t*Bu)$_4$ is located at a similar energy as that of the other MPc molecules. However, fluorescence emission and excitation spectroscopy (Fig. 2b) did not reveal any Q-band emission for NiPc-(*t*Bu)$_4$. As the STML experiments were performed at low temperatures, we also performed optical spectroscopy in solution at 77 K (see Fig. S3), and in PMMA matrices down to $T = 6$ K. The absence of luminescence in all cases confirms that NiPc-(*t*Bu)$_4$ is non-fluorescent, even when optically excited from a vibrational shoulder only 0.28 eV higher in energy than the NiPc-(*t*Bu)$_4$ Q-band.

Next, we show that fluorescence of NiPc can be activated by energy transfer. We built dimers of NiPc with the other MPc molecules, as shown in Fig. 3a, with typical distances (center to center) between the molecules of about $R = 1.45 \pm 0.09$ nm. When the tip was positioned above the NiPc within the dimer, there was still no detectable luminescence. However, when the tip was parked on the lobe of the adjacent MPc molecule, two distinct resonance peaks appeared in the STML spectra (Fig. 3b). In all cases, the high-energy peak was identical to the Q(0,0) resonance of the respective MPc monomer, while the low-energy peak was always located at a photon energy of 1.86 eV. As this low-energy peak only appeared in the presence of the NiPc molecule, and is independent of the chosen MPc molecule, this is a clear sign that this resonance originates from the NiPc molecule and corresponds to its Q-band emission, i.e. $E(Q_{NiPc}) = 1.86$ eV. NiPc was excited into its $S_1$ state via a RET process from the MPc and then radiatively decayed to the ground state. In this scenario, the individual MPc molecules serve as donors exhibiting a higher exciton energy, and the NiPc is the acceptor, i.e., an exciton can be transferred from the MPc to NiPc, but not the other way around.[36-38]

To verify that the dimers were not electronically hybridized, namely strongly bonding, we performed a series of STS spectra along a line across the dimer. Fig. 3c shows an example for a PdPc-NiPc dimer (see Fig. S5 for a corresponding data set on a ZnPc-NiPc dimer). We found that in all cases, the PIR and NIR positions were identical to those of the monomers, shown in Fig. 1. In the region where the two molecules are closest to each other, no continuous transition of the PIR and NIR positions occurred, but the d$I$/d$V$ spectra showed the PIR and NIR from both molecules. This coexistence results from a finite tunneling probability into either of the two molecules, when the tip is located in between the two. This is also confirmed in the STM images, where it can be seen that there is a spatial overlap of the orbital features in this region. Hence, the individual molecular orbital structures were not altered in the dimer, compared to isolated monomers, and therefore the resultant peaks in the STML spectra does not result from a hybrid electronic structure. To further confirm that the NiPc emission is due to a RET mechanism, we also acquired STML spectra as a function of intermolecular distance between a ZnPc and a NiPc molecule (see Supporting Information, Section 4 and Fig. S4), and found that the distance-dependent RET efficiency is similar to previous studies.[36,38]

The RET efficiency varies with the position of the tip with the respect to the MPc donor molecule.[36,38] In Fig. 3d, we present STML spectra and extracted Q-band intensities taken at different positions of a PdPc-NiPc dimer. Both the $Q_{PdPc}$ and $Q_{NiPc}$ emissions were strongest when the tip was placed on PdPc, with maximum distance to the NiPc. As the tip moved closer to the NiPc, the $Q_{NiPc}$ emission initially remained constant up until the center of PdPc was reached, then it continuously decreased and eventually vanished as soon as direct tunneling from NiPc was probable. Compared to that, the $Q_{PdPc}$ emission continuously decreased from the edge of the molecule farthest from NiPc until the PdPc center was reached. From there, the intensity recovered slowly but eventually also tended to vanish when tunneling from the NiPc to the tip became dominant.

We note that also homodimers of two NiPc molecules showed no fluorescence in STML (see Fig. S6a). In comparison, homodimers of ZnPc[32] and PdPc (Fig. S6b) showed a redshift and sharpening of the Q-band emission, which is an expected effect of coherent excitonic coupling.[25,34] Overall, our observations are in line with previous submolecularly resolved STML studies of Pc-based donor-acceptor dimers, confirming our interpretation that the fluorescence of NiPc is enabled by RET from one of the MPc donor molecules (M = Zn, Pd or Pt).

**Discussion**

The activation of NiPc fluorescence via RET can be understood by reviewing the many-body energy diagram of NiPc and considering the various pathways that lead to its excitation. Fig. 4a shows the potential energy of the most relevant NiPc states as a function of the reaction coordinate, which is mostly the Ni–$N_p$ bond distance between the central Ni atom at its four neighboring isoindole N atoms in the center of the Pc macrocycle (see below).[12] Our observations can be rationalized by assuming a small energy barrier between the vibrationally cooled $S_1$ and the (d,d$^*$) state, which has to be overcome to activate the ISC, e.g. by exciting molecular vibrations. In any photoluminescence experiment, an excitation photon energy must be chosen that is larger than the fluorescence energies. Hence, there is sufficient excess energy to overcome this activation barrier. Thus, depending on the energy of the absorbed photon, a vibrationally excited state within the $S_1$ electronically excited state is reached either directly or via excitation into a higher electronically excited state (e.g. $S_2$) followed by relaxation into a vibrationally hot $S_1$ state.[8,10,11] The ultrafast ISC into the $(d,d^*)$ state is possible when the vibrationally excited $S_1$ level is above the activation barrier, leading to radiationless deactivation. This scenario can also explain why optical spectroscopy experiments even at low temperatures still lead to absence of NiPc fluorescence (see Supporting Information, section 3).

In STML experiments of isolated NiPc molecules, the excitation is mostly preceded by an electron tunneling into (or out of) the molecule, leading to an anionic (or cationic) doublet state, which is drawn schematically in Fig. 4a (red arrows).[27,32] In cases where the doublet ground state $D_0$ has a lower energy than the $S_1$ state, the molecule can only go back to $S_0$ by a nonradiative discharging process.[32] In NiPc, both the NIR and PIR threshold magnitudes are accessed at voltages where $eV_s < E(Q_{NiPc})$, hence both anionic and cationic $D_0$ levels are lower than $S_1$. A similar situation was previously found for other non-luminescent Pc molecules.[40,41] The $S_1$ state may still be accessible by applying larger voltages that allow tunneling out of lower-lying occupied (or into higher-lying unoccupied) orbitals. This excites the molecule into a higher doublet state $(D_x)$ and permits access to $S_1$ upon discharging.[32] We have confirmed this by $dI/dV$ and STML spectroscopy down to voltages of $V_s = -3$ V (see Fig. S1), and there is still no detectable NiPc emission despite fulfilling $E(D_x) > E(S_1)$ (see also Section 2 of the Supporting Information). However, there is usually an energy mismatch, and therefore, the discharging will end in a vibrationally hot $S_1$ state. In the case of NiPc, this again activates ultrafast ISC into the dark $(d,d^*)$ state.

Another known excitation channel in STML is plasmon-induced excitation. Molecules can be excited remotely with the tip displaced laterally by a few nanometers. This way, no direct tunneling through the NiPc occurs, but the nanocavity plasmons (NCP) can still couple to and excite the molecule to $S_1$. We conducted such an experiment, but we did not observe any plasmon-induced luminescence for NiPc (see Fig. S7). Obviously, plasmon-exciton coupling can also excite the molecule into a vibrationally hot $S_1$ state (blue arrow in Fig. 4a), hence overcoming the ISC activation barrier and deactivating NiPc emission.

To activate fluorescence in NiPc, it is important to provide an excitation channel into an $S_1$ level that lies below the activation barrier for ISC. This is possible via RET, as shown schematically in Fig. 4b. In a MPc-NiPc dimer, when the tip is positioned above the MPc, the initial excitation via a transiently charged state only occurs in the MPc, as no direct electron tunneling via the NiPc happens. This also results in a vibrationally hot $S_1$ state. However, for the MPc molecules used here (M = Zn, Pd, Pt), there is no lower-lying (d,d$^*$) state. Therefore, the molecule cools into the lowest vibrational $S_1$ level, following Kasha's rule.[34] From there, either radiative decay into the MPc $S_0$ state or a RET process to the neighboring NiPc occurs. As the Q(0,0) energies of all MPc molecules are very close to that of NiPc, the RET leads to a NiPc $S_1$ level that is below the ISC activation barrier. Importantly, the direct optical (or RET) excitation of (d-d$^*$) states is parity forbidden (Laporte's rule).

To rationalize the conceptual potential energy diagram used in Fig. 4, we performed time-dependent density functional theory (TDDFT) calculations (see Supporting Information, section 7). A comparison of the optimized molecular structure in the $S_1$ state as well as various possible (d,d$^*$) states revealed larger Ni–N$_p$ bond distances in case of the latter. Hence, we can identify the "reaction coordinate" axis, which was not defined in a previous theory work,[12] as the Ni–N$_p$ bond length, as shown in Fig. 4. A calculation of the ($\pi,\pi^*$) as well as various singlet and triplet (d,d$^*$) excited-state energies as a function of Ni–N distance (Fig. S8 and S9) also confirms the three main features of our schematic potential diagram in Fig. 4b: (1) the equilibrium Ni–N distance in the $(d,d^*)$ states is larger than in the $(\pi,\pi^*)$ state; (2) almost all of the calculated $(d,d^*)$ states in equilibrium are at lower energy than the $(\pi,\pi^*)$ state; (3) $(d,d^*)$ potential curves intersect with the $(\pi,\pi^*)$ potential, rationalizing an activation barrier that is roughly given by the energy difference between the intersection and the minimum of the $(\pi,\pi^*)$ state. Unfortunately, TDDFT does not allow us to reliably quantify the activation barrier with the degree of accuracy required here. While the above three features are qualitatively robust in all calculations, the quantitative values of potential minima and intersection points are very sensitive to the functional used. Besides, the gas-phase calculations did not include the influence of the surface, which we expect to slightly steepen the potential curves due to the impact on the Ni-N breathing mode.

Finally, we discuss the magnitude of the ISC activation barrier. Our STML experiments showed the largest difference to the $Q_{NiPc}$ emission energy for $Q_{PtPc}$ with $\Delta E$ = 78 meV (Fig. 3b). As RET was still observed, this defines a lower boundary of the activation barrier, $\Delta E_{ISC} \geq 0.08$ eV. We note that absorption resonances are usually at slightly higher energy than emission resonances, known as Stokes shift (see Fig. 2 for Stokes shifts of MPc molecules in solution at room temperature). However, measurements of ZnPc in a cryogenic matrix revealed no detectable Stokes shift,[42] which is why the impact can likely be neglected here. Looking at the results from our calculations, changing the Ni-N bond distance requires a vibrational excitation of about 0.18 eV corresponding to the frequency of 1410.8 cm$^{-1}$ of the Ni-N breathing mode. The intersection point of the potentials is obviously within the uncertainty range of TDDFT, which is typically about 0.3 eV.[43,44] We hence assume this value to be an upper boundary for the ISC activation barrier. This can be confirmed comparing the smallest photon energy used in our photoluminescence experiments (see Supporting Information, section 3) with the NiPc

Q-band energy. From this, we can assume an upper boundary of the ISC activation barrier of $\Delta E_{ISC} \leq 0.28$ eV. These estimates indicate that the excitation energy must be tuned within a relatively narrow range to enable NiPc fluorescence.

A further quantification of the ISC activation barrier would require to use donor molecules with even larger mismatch of the Q-bands, to see at which point it becomes large enough to induce ISC. Another possibility might be an experiment using a carefully tuned tip for remote plasmon-induced excitation. If voltages are applied that excite plasmons with energies $E < E(Q_{NiPc}) + \Delta E_{ISC}$, then plasmon-induced molecular luminescence might be activated. However, this quantum cutoff energy also dramatically cuts off the NCP intensity at $E(Q_{NiPc})$, and finding experimental evidence for NiPc fluorescence under such conditions may prove extremely difficult.

**Conclusion**

In summary, we showed that fluorescence from NiPc molecules can be activated at low temperatures via resonant energy transfer from neighboring MPc molecules. We rationalize this observation by an activation barrier on the order of 0.1 eV for the rapid intersystem crossing from the normally emissive $S_1$ to the dark (d,d*) excited state. In most experimental situations, this barrier is readily overcome by vibrational excitations of the molecule influencing the Ni-$N_p$ bond distance. However, if the $S_1$ energy of neighboring MPc molecules is only slightly above that of NiPc, RET can occur without overcoming the ISC activation barrier. This enables radiative decay from $S_1$ to the $S_0$ ground state, and hence Q-band emission of NiPc. This strategy should also work for other compounds that possess rapid deactivation of luminescence, e.g., other open-shell dark metallophthalocyanines such as VPc, FePc and CoPc, or corresponding metal porphyrins. Beyond that, more systematic studies of the conditions under which RET enables molecular luminescence may shed more light on the underlying mechanisms of intermolecular energy transfer, with relevance in photosynthesis and photovoltaics.[45,46]

Our findings encourage the exploration of rigidified local environments at low temperatures, which permit the directed excitation of the vibrationally "cold" electronic excited states to enable the rather rare emission from dark 3$d$-metal complexes. In this context, it would be interesting to develop ultrafast transient absorption spectroscopy toward low temperature studies of molecules in frozen matrices or on solid-state substrates. This may allow to further detail out and quantify the NiPc excited states energy diagram.[12] Furthermore, a deeper understanding of the nature and role of molecular energy transfer in electroluminescence enables to establish intermolecular design strategies toward efficient lighting applications, in addition to the already established strategy to tweak the intramolecular structure.[23] Thus, the use of rare metals such as Ir or Pt could be replaced by more abundant exemplars involving Ni(II) complexes.[5]

**Acknowledgements**

We thank Jascha Repp for support and stimulating discussions. We also acknowledge Abhijnan Chatterjee for support during some of the experiments. This publication is part of the project OCENW.M20 financed by the Dutch Research Council (NWO). T.-C. H acknowledges support from the ERC Synergy Grant MolDAM (no. 951519). B. K. acknowledges the NWO-VENI project 016.Veni.192.168. N.L.D. and C.A.S. gratefully acknowledge funding from the Deutsche Forschungsgemeinschaft (DFG, German Research Foundation)–project-ID

## Associated Content

**Supporting Information:** Experimental details, d$I$/d$V$ and bias dependence of STML spectra, temperature-dependent photoluminescence spectra of NiPc, distance dependence of energy transfer between ZnPc and NiPc, STS on ZnPc-NiPc dimers, STML on homodimers, spatial dependence of STML, theoretical details and analysis, NiPc vs. HPc$^-$ comparison.

## Author Information

**Corresponding Author**

> **Daniel Wegner** – Institute for Molecules and Materials, Radboud University, 6500 GL Nijmegen, The Netherlands; orcid.org/0000-0002-1625-2830; Email: d.wegner@science.ru.nl

**Authors**

> **Tzu-Chao Hung** – Institute for Molecules and Materials, Radboud University, 6500 GL Nijmegen, The Netherlands; Institute for Experimental and Applied Physics, University of Regensburg, 93040 Regensburg, Germany; orcid.org/0000-0001-5583-6612
>
> **Yokari Godinez-Loyola** – Institut für Anorganische und Analytische Chemie, University of Münster, 48149 Münster, Germany; CeNTech, University of Münster, 48149 Münster, Germany; orcid.org/0000-0002-3390-6250
>
> **Manuel Steinbrecher** – Institute for Molecules and Materials, Radboud University, 6500 GL Nijmegen, The Netherlands; orcid.org/0000-0003-3250-402X

**Brian Kiraly** – Institute for Molecules and Materials, Radboud University, 6500 GL Nijmegen, The Netherlands; orcid.org/0000-0002-4666-9290

**Alexander A. Khajetoorians** – Institute for Molecules and Materials, Radboud University, 6500 GL Nijmegen, The Netherlands; orcid.org/0000-0002-6685-9307

**Nikos L. Doltsinis** – Institut für Festkörpertheorie and Center for Multiscale Theory and Computation, University of Münster, 48149 Münster, Germany; orcid.org/0000-0003-0669-5854

**Cristian A. Strassert** – Institut für Anorganische und Analytische Chemie, University of Münster, 48149 Münster, Germany; CeNTech, University of Münster, 48149 Münster, Germany; Cells in Motion Interfaculty Centre (CiMIC) and Center for Soft Nanoscience (SoN), University of Münster, 48149 Münster, Germany; orcid.org/0000-0002-1964-0169

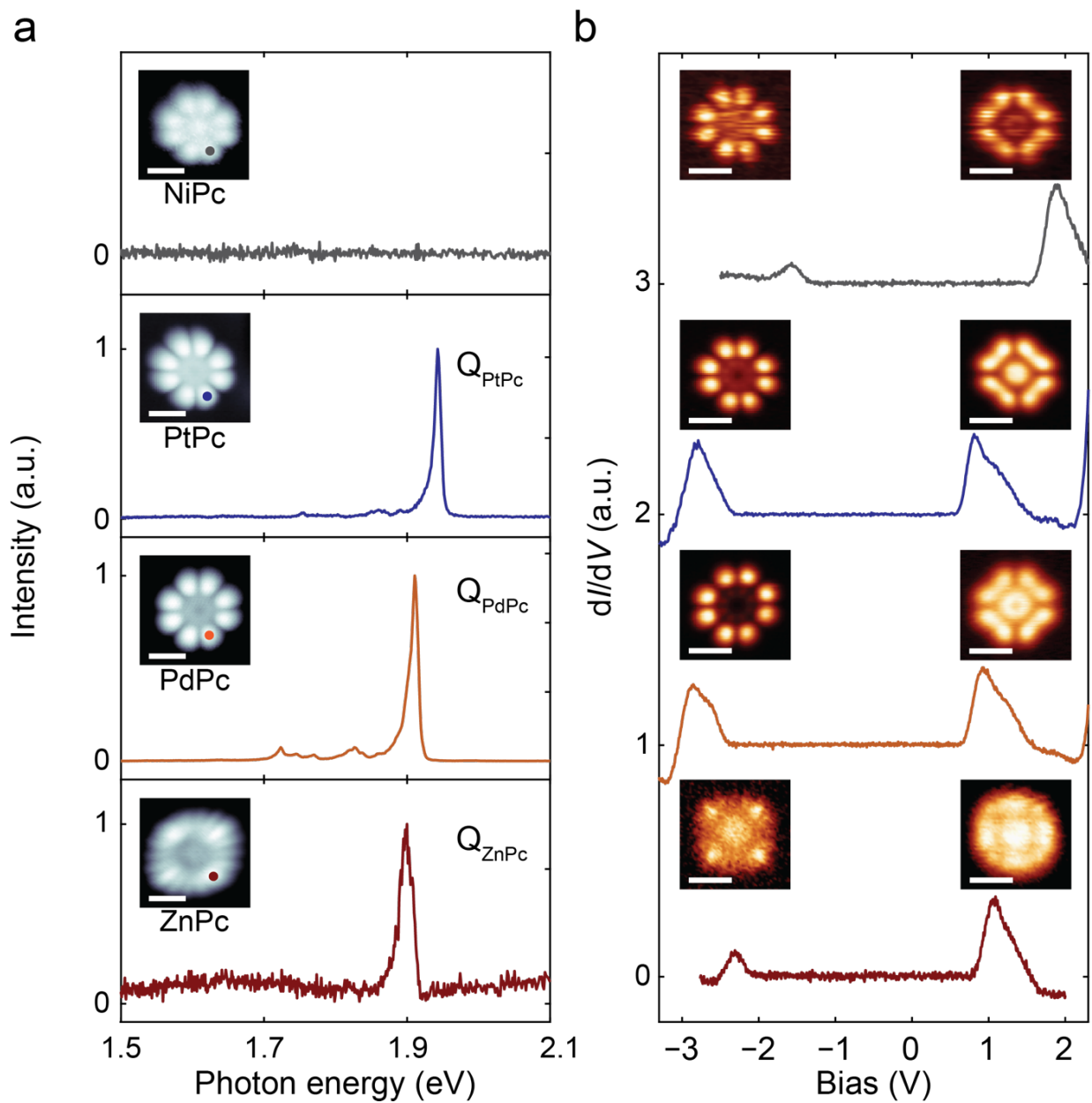

**Figure 1. STM-based optical and electronic characterization of individual MPc molecules.** (a) STML spectra of NiPc, PtPc, PdPc and ZnPc on 3 ML NaCl/Ag(111). While NiPc did not show any fluorescence, the other MPc's exhibited the well-known Q(0,0) fluorescence. The insets show constant-current STM images of the respective MPc molecules, marking positions where the tip was parked for the STML measurements. (b) Differential conductance spectra of the MPc molecules shown in (a), showing the positions of the HOMO and LUMO as found in STS. The insets to the left (right) show the spatial d$I$/d$V$ maps of the HOMO (LUMO) orbitals for each molecule respectively, confirming the assignment. (Parameters used for the data acquisition are given in the Supporting Information, section 1.)

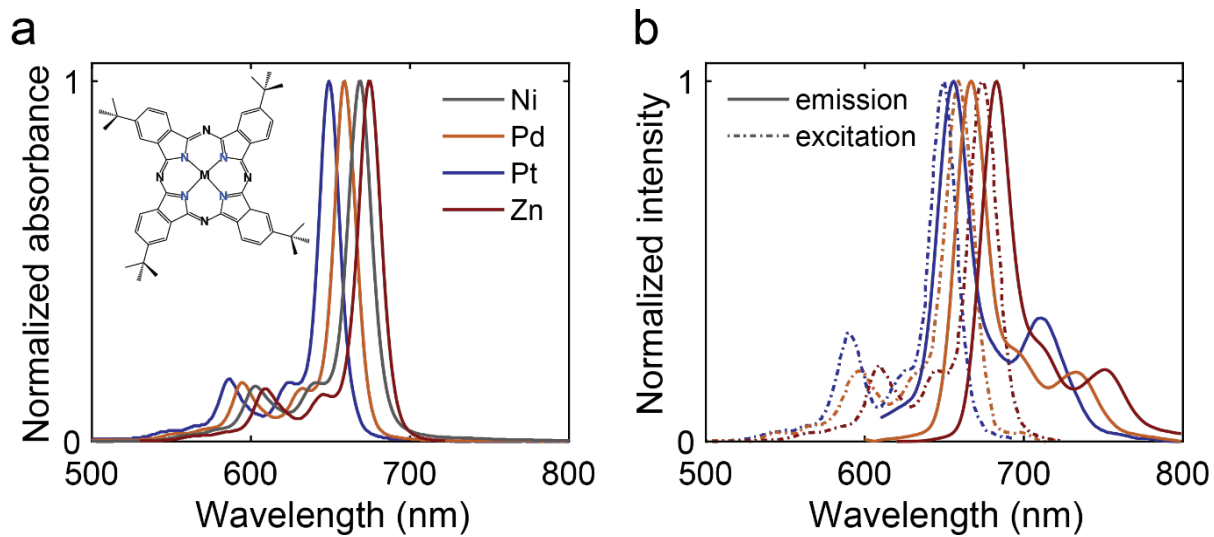

**Figure 2: Optical properties of phthalocyanines in solution.** (a) Absorption spectra of NiPc-(*t*Bu)$_4$ and MPc-(*t*Bu)$_4$ (M = Zn, Pd and Pt) in a 2Me-THF:toluene solution taken at room temperature. The *tert*-butyl groups have been added for solubility reasons but should not significantly affect the optical properties. (b) Fluorescence emission (solid line) and excitation (dashed line) spectra of MPc-(*t*Bu)$_4$ at room temperature. For NiPc-(*t*Bu)$_4$, no emission was found.

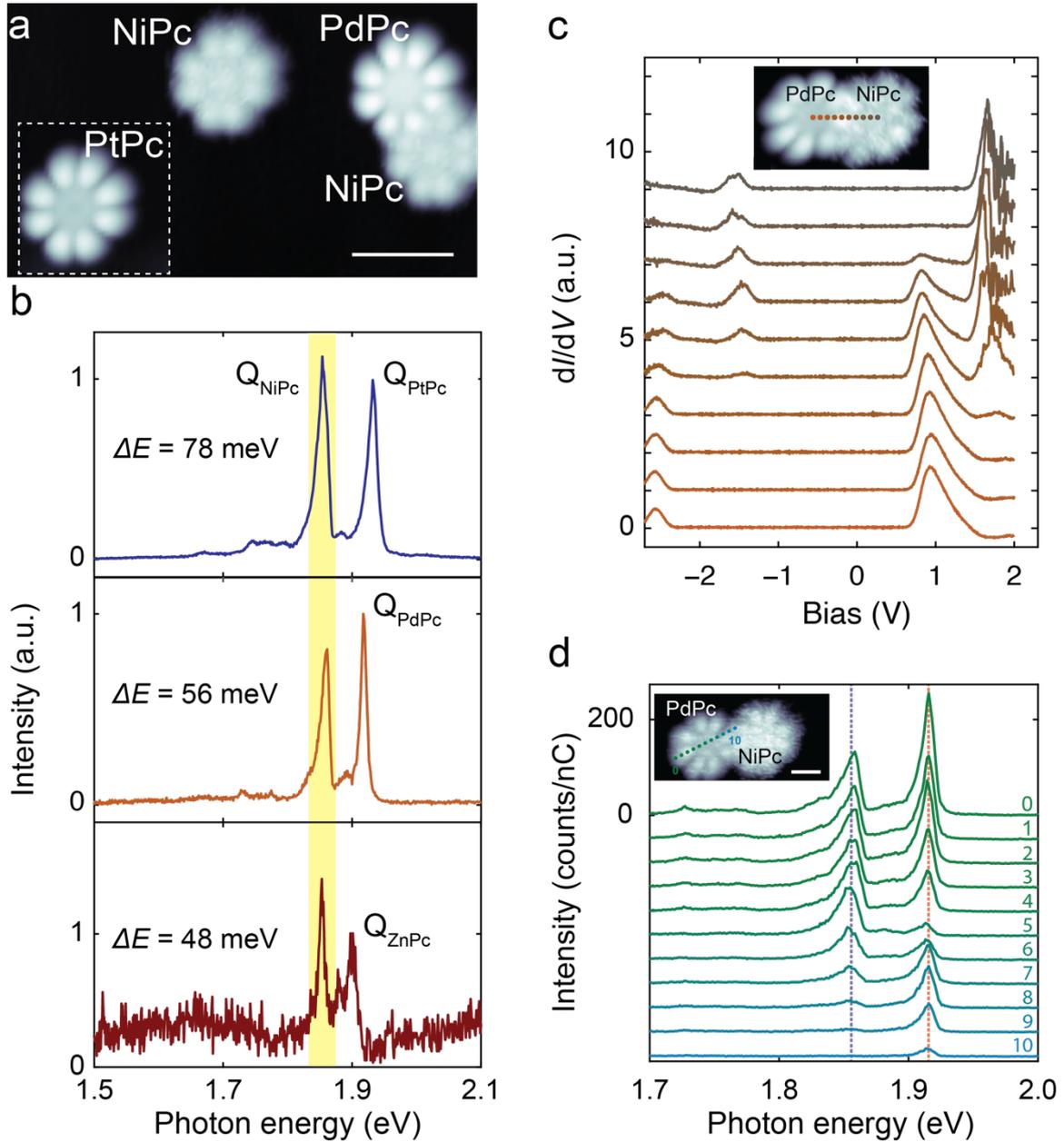

**Figure 3. Resonant energy transfer and fluorescence of NiPc in MPc dimers.** (a) Constant-current STM image showing MPc monomers and a PdPc-NiPc dimer with center-to-center distance $R \approx 1.43$ nm (scalebar: 2 nm; stabilization parameters: $V_s = -2.5$ V, $I_t = 10$ pA). (b) STML spectrum of MPc-NiPc dimer with M = Pt (top), Pd (center) and Zn (bottom), respectively. The tip was always parked above the MPc. In addition to the Q-band emission of the respective MPc, a second sharp emission peak $Q_{NiPc}$ was observed, highlighted by the yellow background ($I_t = 100$ pA, $V_s = -2.6$ V, $t = 120$ s). (c) Set of d$I$/d$V$ spectra taken across a PdPc-NiPc dimer (see inset), revealing that the individual molecular orbitals are not altered compared to isolated monomers (feedback loop opened at the center of PdPc with $V_s = -2.8$ V, $I_t = 50$ pA). (d) Set of STML spectra taken across a PdPc-NiPc dimer with $R \approx 1.85$ nm. $Q_{NiPc}$ emission was strongest when the tip was parked on the PdPc with maximum distance to the NiPc, and it vanished as soon as tunneling into NiPc was possible ($I_t = 100$ pA, $V_s = -2.5$ V, t = 120 s).

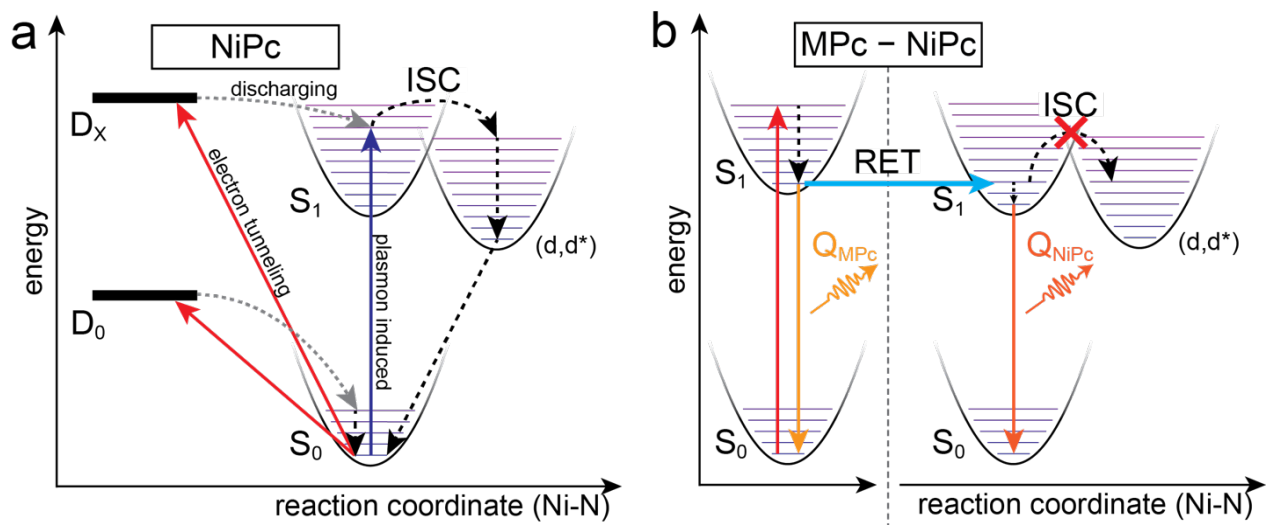

**Figure 4. Energy diagram of NiPc excitations and energy transfer.** (a) Schematic potential energy diagram of NiPc monomer showing tunneling- (red arrows) or plasmon-induced (blue) excitation channels that lead to ISC and radiationless deactivation. (b) Schematic potential energy diagram of MPc-NiPc dimer showing how RET activates radiative decay from the $S_1$ state by preventing ISC.

Supporting Information

# Activating the fluorescence of a Ni(II) complex by energy transfer

Tzu-Chao Hung[1,2], Yokari Godinez-Loyola[3,4], Manuel Steinbrecher[1], Brian Kiraly[1], Alexander A. Khajetoorians[1], Nikos L. Doltsinis[5], Cristian A. Strassert[3,4,6], and Daniel Wegner[1*]

**Affiliations:**

[1] Institute for Molecules and Materials, Radboud University, 6500 GL Nijmegen, The Netherlands

[2] Institute for Experimental and Applied Physics, University of Regensburg, 93040 Regensburg, Germany

[3] Institut für Anorganische und Analytische Chemie, University of Münster, 48149 Münster, Germany

[4] Center for Nanotechnology (CeNTech), University of Münster, 48149 Münster, Germany

[5] Institut für Festkörpertheorie and Center for Multiscale Theory and Computation, University of Münster, 48149 Münster, Germany

[6] Cells in Motion Interfaculty Centre (CiMIC) and Center for Soft Nanoscience (SoN), University of Münster, 48149 Münster, Germany

* Email: d.wegner@science.ru.nl

**The supporting information includes:**

1. **Experimental details**
2. **d$I$/d$V$ spectra and bias dependence of STML**
3. **Temperature-dependent photoluminescence spectra of NiPc**
4. **Distance dependence of energy transfer between ZnPc and NiPc**
5. **STS on ZnPc-NiPc dimers**
6. **STML on homodimers**
7. **Spatial dependence of STML**
8. **Theoretical details and analysis**
9. **NiPc vs. HPc⁻**

# 1. Experimental details

ZnPc was purchased from Sigma-Aldrich. NiPc PdPc and PtPc as well as NiPc-(*t*Bu)$_4$, ZnPc-(*t*Bu)$_4$, PdPc-(*t*Bu)$_4$ and PtPc-(*t*Bu)$_4$ were kindly supplied by PorphyChem (Dijon, France).

The STM experiments were carried out in a commercial Omicron ultra-high vacuum (UHV) low-temperature STM system operated at $T = 4.5$ K with a base pressure below $1\times10^{-10}$ mbar.[1] We used a silver bulk tip, which was electrochemically etched in a mixture of perchloric acid and methanol (ratio 1:4),[2,3] and further treated in UHV by field emission and controlled indentation on a Ag(111) surface. The Ag(111) single crystal (MaTeck) was cleaned by multiple cycles of sputtering and annealing followed by NaCl deposition while the Ag(111) surface was kept at room temperature. The NiPc and MPc molecules were successively sublimated from a Knudsen cell evaporator and deposited on the surface held at $T < 6$ K inside the STM. We note that the evaporator was always thoroughly degassed in UHV by heating to the target temperature for tens of minutes (in a UHV space decoupled from the main UHV chambers), in order to reduce contaminations in the source (see also Section 9).

All STML spectra were acquired using a 150 grooves/mm grating. Further details of the optical setup can be found in a previous publication.[1] PdPc-NiPc and ZnPc-NiPc dimers were created by lateral atomic manipulation of the PdPc and ZnPc molecules, respectively, following recipes reported in the literature.[4,5] We note that we were not able to reliably manipulate NiPc molecules. The PtPc-NiPc dimer whose STML spectrum is shown in Fig. 2 was found self-assembled on the surface.

The STML spectra presented in Fig. 1a were acquired in constant-current mode using the following stabilization voltage ($V_s$), tunnel current ($I_t$) and acquisition time $t$: $V_s = -2.5$ V, $I_t = 100$ pA, $t = 300$ s (NiPc); $V_s = -2.6$ V, $I_t = 50$ pA, $t = 120$ s (PtPc); $V_s = -2.5$ V, $I_t = 100$ pA, $t = 120$ s (PdPc and ZnPc). The inset constant-current topography images were acquired using $V_s = -2.6$ V, $I_t = 10$ pA (NiPc and PtPc) and $V_s = -2.5$ V, $I_t = 10$ pA (PdPc and ZnPc).

The d$I$/d$V$ spectra shown in Fig. 1b were acquired with the feedback loop opened at the center of the molecule using: $V_s = +2.5$ V, $I_t = 200$ pA (NiPc); $V_s = -3.5$ V, $I_t = 100$ pA (PtPc); $V_s = -3.0$ V, $I_t = 100$ pA (PdPc); $V_s = -2.8$ V, $I_t = 50$ pA (ZnPc). The inset constant-height d$I$/d$V$ maps were taken with the tip stabilized above the center of each molecule prior to opening the feedback loop, at $I_t = 100$ pA and $V_s$ set close to the respective maximum of the HOMO/LUMO peak, which was $-1.5$ V/$+1.5$ V (NiPc), $-2.5$ V/$+0.7$ V (PtPc), $-2.4$ V/$+0.85$ V (PdPc), and $-2.25$ V/$+0.8$ V (ZnPc), respectively.

For the photophysical characterization, all solvents used were of spectroscopic grade (Uvasol®). Absorption spectra were measured with a Shimadzu UV-3600 I plus UV-VIS-NIR spectrophotometer. Steady-state excitation and emission spectra were recorded on a FluoTime300 spectrometer from PicoQuant equipped with a 300 W ozone-free Xe lamp (250-900 nm), a 10 W Xe flash-lamp (250-900 nm, pulse width ca. 1 μs) with repetition rates of 0.1 – 300 Hz, a double-grating excitation monochromator (Czerny-Turner type, grating with 1200 g/mm, blaze wavelength: 300 nm), two double-grating emission monochromators (Czerny-Turner, selectable gratings blazed at 500 nm with 2.7 nm/mm dispersion and 1200 grooves/mm, or blazed at 1200 nm with 5.4 nm/mm dispersion and 600 grooves/mm) with adjustable slit width between 25 μm and 7 mm, Glan-Thompson polarizers for excitation (Xe-lamps) and emission (after the sample). Different sample holders (Peltier-cooled four-position cuvette sample holder ranging from -15 to 110 °C and round cuvette sample holder), along with two detectors, namely a PMA Hybrid-07 (transit time spread FWHM < 50 ps, 200 – 850 nm) and an H10330C-45-C3 NIR detector (transit time spread FWHM 0.4 ns, 950-1700 nm) from Hamamatsu were used. Emission and excitation spectra were corrected for source intensity (lamp and grating) by standard correction curves.

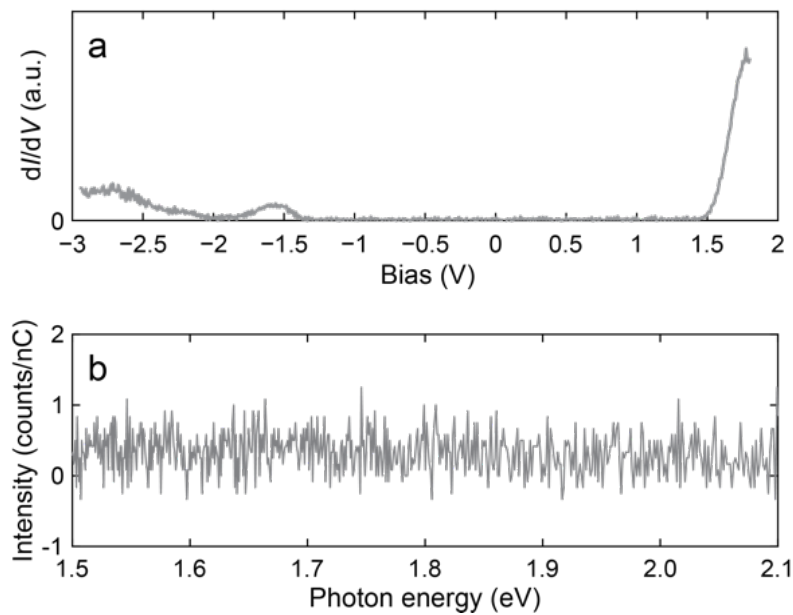

**Figure S1. Large negative bias range STS and STML spectra.** (a) d$I$/d$V$ spectrum on the center of NiPc adsorbed on 3 ML NaCl/Ag(111), extended down to –3 V (feedback loop opened at NiPc center with $V_s = -2.6$ V, $I_t = 100$ pA). An additional peak can be found at –2.7 V, and a shoulder is visible at about –2.3 V. (b) STML spectrum acquired on NiPc on 3 ML NaCl/Ag(111) at large negative sample bias ($V_s = -3$ V, $I_t = 100$ pA, $t = 120$ s).

## 2. d$I$/d$V$ spectra and bias dependence of STML

An obvious difference of NiPc over the other MPc molecules is a relatively large shift of the PIR and NIR by about 1 V, and we would like to provide a hypothesis for this. In the energy range of the NiPc frontier orbitals, several d orbitals can be found from theory (see also Section 8 and Table S2).[6] Especially the $d_\pi$ orbitals (i.e. $d_{xz}$ and $d_{yz}$) can have a relatively strong interaction with the ligand-π orbitals, which could shift the positions of the latter. PdPc and PtPc have a much larger ligand field splitting, owing to the 4d and 5d character (as opposed to 3d for NiPc). Therefore, their $d_\pi$ orbitals are lying deeper in energy and do not influence the frontier orbitals so much. As ZnPc has a completely filled d-shell, its d orbitals are also at lower energies. To confirm this hypothesis, d$I$/d$V$ spectra of other 3d transition metal phthalocyanines, e.g. MnPc, FePc and CoPc should be compared with each other on NaCl/Ag(111). In this context, we note that a comparison of MnPc with MgPc molecules on NaCl/Cu(111) showed a comparably strong shift in the d$I$/d$V$ spectra.[7]

Fig. S1a shows a d$I$/d$V$ spectrum acquired on NiPc extended down to −3 V, which indicates the presence of at least two occupied molecular orbitals below −2 V. Hence, tunneling an electron out of one of those orbitals excites the NiPc molecule into a transiently charged (i.e. NiPc$^+$) doublet state $D_x$ that fulfills the condition $E(D_x) > E(S_1)$. This shows that already the STML spectra taken at −2.5 V (Fig. 1a) and −2.6 V (Fig. S4 and S5) provide evidence that NiPc in non-emissive in STML. To confirm this, we provide STML spectra taken at −3 V (Fig. S1b). Again, the NiPc is non-emissive even under such extreme condition.

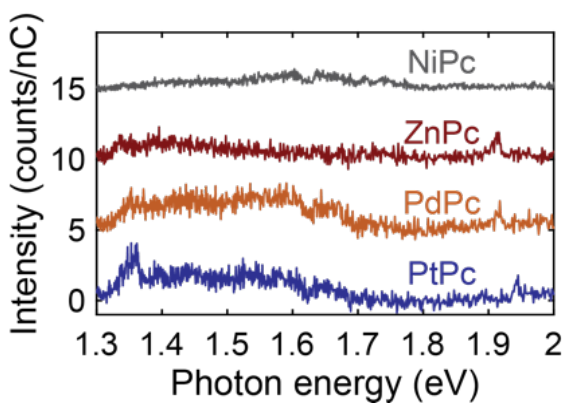

**Figure S2. Bias dependence of STML spectra.** STML spectra acquired on 3 ML NaCl/Ag(111) at positive sample bias of NiPc (grey, $V_s$ = 2.5 V, $I_t$ = 100 pA, $t$ = 300 s.), ZnPc (red, $V_s$ = 2.5 V, $I_t$ = 50 pA, $t$ = 300 s.), PdPc (orange, $V_s$ = 2.6 V, $I_t$ = 100 pA, $t$ = 120 s.), and PtPc (blue, $V_s$ = 2.5 V, $I_t$ = 100 pA, $t$ = 120 s.). The spectra are offset for clarity.

To verify that individual NiPc molecules on 3 ML NaCl/Ag(111) do not show any Q-band emission, irrespective of the tunneling conditions, we also acquired STML spectra at positive sample bias. As shown in Fig. S2, no STML intensity was found around $E(Q_{NiPc})$ = 1.86 eV. The spectrum only contained weak and spectrally broad features stemming from radiative NCP decay. We note, however, that at positive bias it is not clear if a magnitude of 2.5 V is sufficient to access an orbital that excites NiPc into a doublet state that is higher in energy than the $S_1$ state. If we were to assume a rigid shift of all ligand orbitals, a voltage of more than 3 V would need to be applied, based on previous studies of ZnPc.[8] However, tunneling at such high positive voltages became unstable.

In comparison, STML spectra of MPc molecules (M = Zn, Pd, Pt) exhibit the respective Q-band emissions at the same energies as was the case for negative sample bias (Fig. 1a). The STML yield at positive sample bias is at least an order of magnitude smaller, explaining the noisier spectra in Fig. S2. From a previous study focusing on ZnPc, we found indications that there is no plasmonic enhancement of STML at positive bias.[8] The spectra shown here indicate that this may also be the case for the other MPc molecules. In addition to the Q-band emission of the neutral molecule, another faint peak can be observed for all molecules at around 1.35 eV. Previously, we showed that for ZnPc, this can be assigned to the radiative decay of a trion state, i.e., excitonic decay from a transiently charged ZnPc⁻ molecule.[8] The STML spectra shown in Fig. S2 were acquired with a tip that was not optimized to enhance this spectral feature, which is why we only observed a faint peak at 1.33 eV. While the STML spectrum of PdPc is less clear, PtPc shows a spectral feature at similar energy (ca. 1.36 eV), indicating that this may stem from the radiative decay of the transiently charged PtPc⁻. Again, the STML spectrum of NiPc is featureless also in this energy range.

## 3. Temperature-dependent photoluminescence spectra of NiPc

To confirm that NiPc does not show any fluorescence, we performed luminescence spectroscopy of various samples and at different temperatures. Fig. S3 summarizes results for NiPc-($t$Bu)$_4$ in a fluid 2Me-THF:toluene 1:1 mixture solution at room temperature as well as at

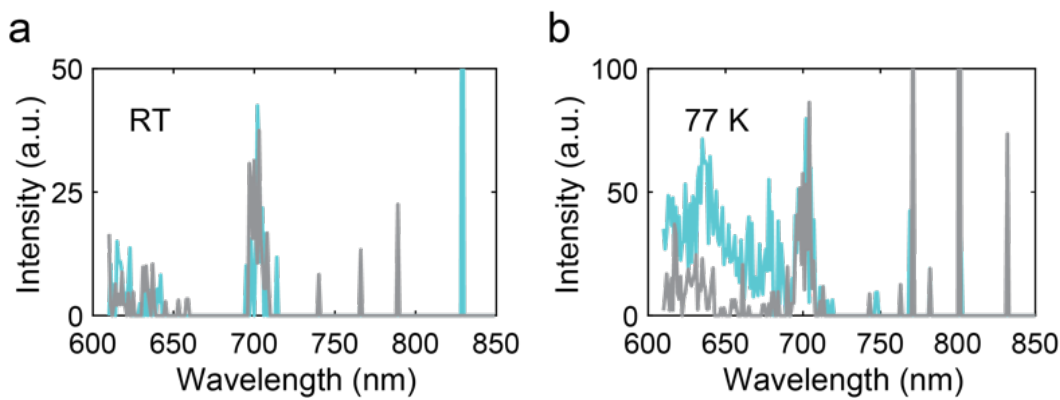

**Figure S3. Absence of NiPc-(*t*Bu)$_4$ photoluminescence.** (a) Luminescence spectrum of NiPc-(*t*Bu)$_4$ in 2Me-THF:toluene in a 1:1 ratio (grey) taken at room temperature. For comparison, the luminescence spectrum of the pure solvent without NiPc-(*t*Bu)$_4$ is also shown (cyan). (b) Same as (a) but taken at 77 K. For all spectra, an excitation wavelength of 580 nm was used, the detector and excitation bandwidths were set to 4 nm, and no filter was used.

77 K using an excitation wavelength of 580 nm (i.e., a photon energy of 2.138 eV). We used this NiPc derivative to increase solubility, noting that the *t*Bu groups should not have any significant impact on the optical properties (especially unwanted aggregation phenomena). For comparison, also the spectra of the bare solvent are shown. To be able to roughly relate the spectra with those shown in Fig. 2b, we used the same excitation and detector settings as for the measurements of PtPc-(*t*Bu)$_4$, which yielded peak intensities that were more than 2000 times higher than those shown in Fig. S3.

The only spectral feature that can be observed is a very faint peak located at ~700 nm. However, it is also observed in the pure solvent spectra and therefore cannot be assigned to an emission from NiPc-(*t*Bu)$_4$. Based on the absorption spectrum shown in Fig. 2 and considering a possible Stokes shift, we would expect a possible Q-band fluorescence peak of NiPc-(*t*Bu)$_4$ to be located around 670-680 nm with a width of ~20-30 nm. Clearly, there is no such feature visible in any of the spectra. We also carried out optical spectroscopy of NiPc-(*t*Bu)$_4$ (as well as the other MPc derivatives) in a solid state PMMA film, down to a temperature of 6 K. While all other MPc-(*t*Bu)$_4$ samples still show luminescence, NiPc-(*t*Bu)$_4$ still lacks any emission. We note that we repeated measurements using also various smaller excitation wavelengths, down to 350 nm, leading to similar results. To avoid saturation of the detector by scattered light from the excitation source, we did not perform measurements at excitation wavelengths larger than 580 nm.

## 4. Distance dependence of energy transfer between ZnPc and NiPc

We further investigated the nature of the Q$_{NiPc}$ emission and the underlying energy transfer mechanism by acquiring STML spectra as a function of intermolecular distance $R$ between a ZnPc and a NiPc molecule (Fig. S4). Here, the tip was always positioned at the equivalent location on the ZnPc molecule (marked by a red dot in the inset STM images). We observed NiPc emission only at relatively close distances $R < 2$ nm. At the closest distance $R = 1.45\pm0.09$ nm, the RET efficiency[4] was highest with 69±2%, and it quickly decreased with increasing $R$. For $R \geq 2$ nm, we only observed the fluorescence of ZnPc. We note that the Q$_{ZnPc}$ intensity at $R = 2.94$ nm is larger than at $R = 2.16$ nm. This indicates that there may still be a coupling to the NiPc, but any possible RET is not discernible within the noise. This relatively short-ranged

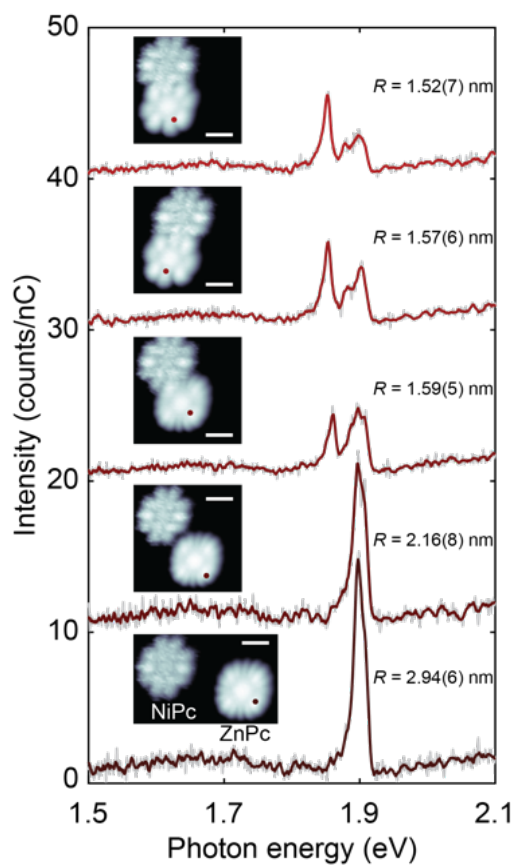

**Figure S4. Distance-dependent RET from ZnPc to NiPc.** STML spectra taken on a ZnPc molecule with different lateral separation $R$ (center to center) from a neighboring NiPc molecule (see inset STM images; scalebar: 1 nm). Below $R = 2$ nm, fluorescence from ZnPc and NiPc was observed, while for larger separations only ZnPc fluorescence was evident. The STML spectra were acquired in constant-current mode with the tip parked on the ZnPc molecule at the position marked by a red dot in the respective inset topography images, using $V_s = -2.5$ V, $I_t = 200$ pA, $t = 120$ s (for all cases where $R < 2$ nm) as well as $V_s = -2.5$ V, $I_t = 100$ pA, $t = 120$ s (for $R > 2$ nm). The inset STM images were taken in constant-current mode using $V_s = -2.5$ V, $I_t = 10$ pA.

RET is in line with previous observations of RET distance dependence based on STML measurements.[4,9]

## 5. STS on ZnPc-NiPc dimers

Figure S5 shows a set of d$I$/d$V$ spectra similar to that shown in Fig. 3c, but for a ZnPc-NiPc dimer. As discussed in the main text, the PIR and NIR positions are identical to those of the monomers, and no continuous transition of the PIR and NIR positions occurs in the interface region between the two molecules. This verifies that the dimers are not electronically hybridized, but the molecules are rather physisorbed next to each other.

## 6. STML on homodimers

In order to test the nature of intermolecular interactions that lead to the NiPc fluorescence, we investigated NiPc homodimers that were found self-assembled on the surface. As can be seen in Fig. S6a, no molecular emission was observed on this NiPc-NiPc dimer, irrespective of the tip position. In comparison, a PdPc-PdPc dimer (Fig. S6b) shows strong fluorescence emission. The peak energy varies between 1.80 and 1.82 eV, depending on the tip position. We note that this energy is red-shifted compared to the fluorescence of the monomer, likely due to coherent intermolecular dipole-dipole interaction.[5] We observed similar behavior also for ZnPc-ZnPc dimers.[8] A simple interpretation of homomolecular chains is that excitons are delocalized across the entire chain, which could also be interpreted in a tight-binding picture as excitons hopping between molecules. This again could be considered a special "resonant" case of RET, where donor and acceptor have the same energy. Obviously, such a scenario still cannot induce fluorescence in NiPc, presumably because the ISC in the initially excited NiPc occurs much faster than the energy transfer of a $(\pi,\pi^*)$ exciton to a neighboring NiPc.

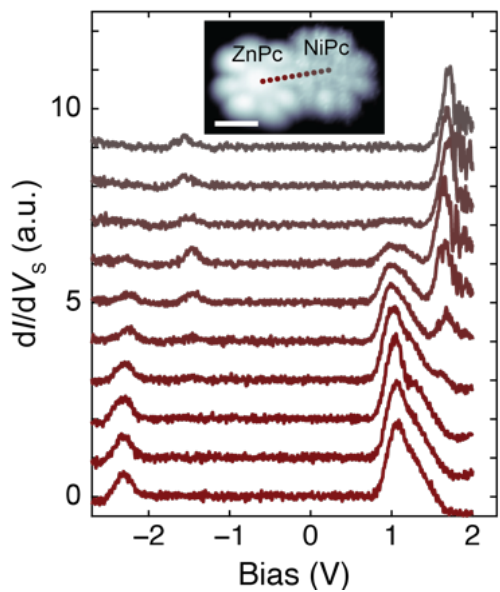

**Figure S5. Electronic structure characterization of ZnPc-NiPc dimer.** Series of constant-height d$I$/d$V$ spectra taken along a line across the dimer (see inset), revealing that the individual molecular orbital structures are not altered in the dimer compared to isolated monomers. Feedback loop was opened at the center of ZnPc molecule with $V_s = -2.8$ V, $I_t = 50$ pA. Inset STM image was acquired at $V_s = -2.5$ V, $I_t = 10$ pA (scalebar: 1 nm).

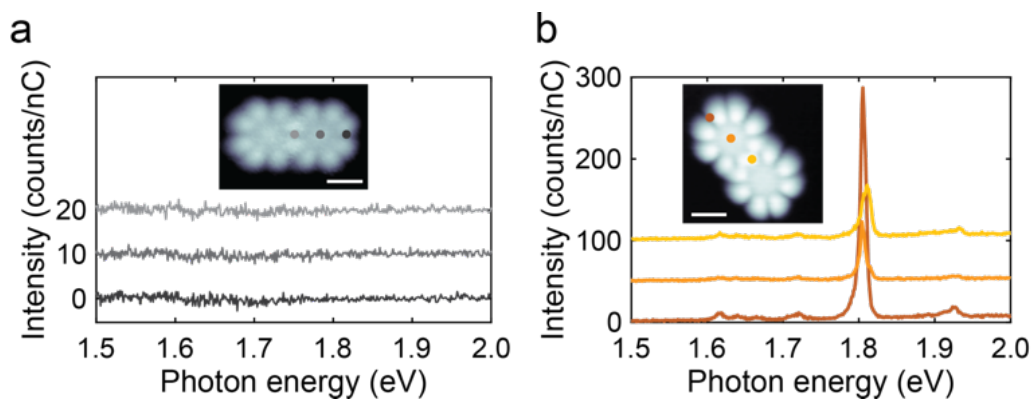

**Figure S6. STML on homodimers.** (a) STML spectra on a NiPc-NiPc dimer at various positions show no indication of any molecular emission. (b) For comparison, STML spectra on a PdPc-PdPc dimer reveal a strong molecular emission at about 1.8 eV, as well as some vibrational side peaks. Inset STM images were acquired at $I_t = 10$ pA, $V_s = -2.6$ V (scalebar: 1 nm). STML acquisition parameters: $V_s = -2.6$ V, $I_t = 10$ pA, $t = 120$ s.

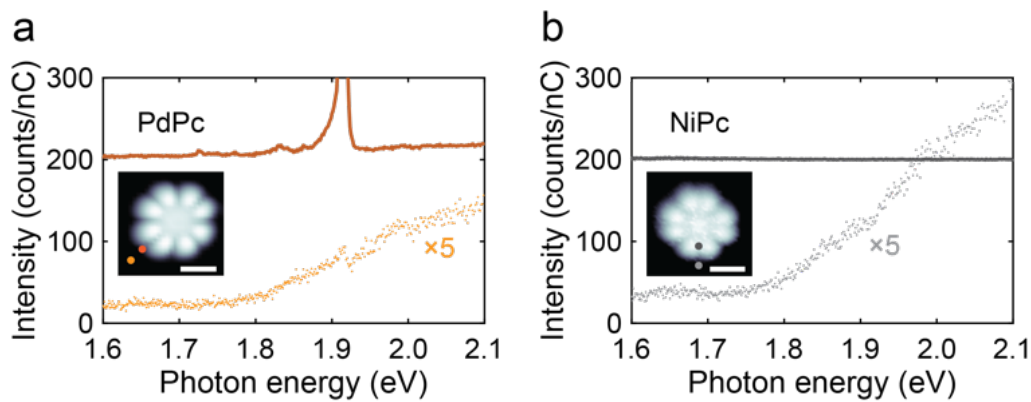

**Figure S7. Plasmon-induced STML.** (a) STML spectrum on (top) and off (bottom) PdPc. In the latter case, a Fano resonance is observed at 1.92 eV when the tip is located close (< 2 nm) to the PdPc but no direct tunneling into the molecules occurs. (b) STML spectrum on (top) and off (bottom) NiPc. For the latter, no Fano resonance is observed, and all observed STML intensity is identical to that of spectra taken far away from any molecule. Inset STM images were acquired at $I_t$ = 10 pA, $V_s$ = −2.6 V (scalebar: 1 nm). STML acquisition parameters: $I_t$ = 100 pA, $V_s$ = −2.6 V, $t$ = 120 s.

## 7. Spatial dependence of STML

In order to test whether molecular fluorescence of NiPc can be activated by plasmon-induced excitation, we compared STML spectra taken on top of the molecule (i.e., while still directly tunneling through it) with those taken when the tip was positioned laterally away from the molecule to prevent direct tunneling. For comparison, we first show results for a PdPc molecule (Fig. S7a). When the tip is positioned at the edge of the macrocycle (i.e. resonant tunneling through the molecule is still possible), the STML spectrum (red) shows an intense peak at 1.91-1.92 eV. We note a slight blueshift compared to the spectrum in Fig. 1 may be due to a reduced Lamb shift.[10] When placing the tip off the molecule (see points marked in the inset STM image), the STML spectrum is dominated by the plasmon resonance of the tip-sample nanocavity (often referred to as nanocavity plasmons (NCPs)), but a faint Fano resonance is visible at 1.92 eV, which is the hallmark of coherent plasmon-exciton coupling and hence leads to faint molecular fluorescence.[11]

In comparison, NiPc STML spectra using the same tip show no indication of molecular fluorescence, irrespective of where the tip is located during the STML acquisition (Fig. S7b). We note that the faint broad feature between 1.8 eV and 1.9 eV cannot be ascribed to the NiPc, as it was also visible in NCP spectra on NaCl/Ag(111) further away from the molecule, using the same tip. We verified that even at applied voltages of $V_s$ = ±3 V, the results remain the same. We therefore conclude that plasmon-induced excitation is not able to activate NiPc fluorescence, for the voltages used. We note that the voltage magnitudes used for the NCP excitation were much larger than $E(Q_{NiPc})$. The observations could be explained by assuming that also higher-energy NCPs can transfer energy to the NiPc and excite it into a vibrationally hot $S_1$ state, leading to ISC.

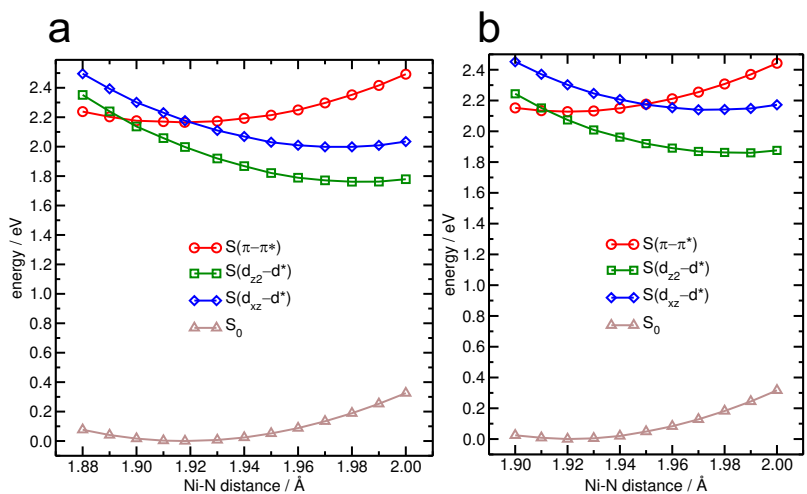

**Figure S8. TDDFT excited-state energy profiles of NiPc for varying Ni-N distances.** (a) Using the PBE0 hybrid functional; (b) using the range-separated hybrid functional CAM-B3LYP. Here, only singlet excited states were considered. For comparison, the optimized Ni-N distance in the ground state was found at 1.918 Å (PBE0) and 1.920 Å (CAM-B3LYP), respectively.

## 8. Theoretical details and analysis

All density functional theory (DFT) calculations were performed using the quantum chemistry package Gaussian 09 Rev. D.01[12] with the PBE0[13] and CAM-B3LYP[14] exchange-correlation functionals and the SDD basis set, which applies an effective core potential for the Pt and Pd atoms[15] and the D95 basis set for H, C, N, F and O atoms.[16] Excited state energy profiles were calculated using time-dependent DFT (TDDFT) linear response theory along the Ni-N reaction coordinate at geometries optimized with all four Ni-N bonds constrained to the same distance. Due to occurring triplet instabilities, additional TDDFT calculations were carried out in the Tamm-Dancoff approximation to obtain more reliable energies of the triplet states relative to the singlet manifold. Molecular orbitals (MOs) were visualized using Avogadro 1.2.0.[17]

From the results of TDDFT, we found that there is a number of *d-d* exited states located at energies below that of the $\pi$-$\pi^*$ singlet state (see also Table S1). The molecular structures optimized in these two types of electronic states mostly differ in the bond length between the central Ni atom and the four neighboring isoindole $N_p$ atoms, hence we can identify the Ni-N bond distance as a suitable reaction coordinate. A corresponding vibrational mode that changes this bond distance in our calculations is the symmetric Ni-N stretch vibration at 1411 cm$^{-1}$ ≈ 175 meV. To explore the reasons why fluorescence only occurs upon resonant excitation of the lowest excited $\pi$-$\pi^*$ singlet state, we scanned the potential energies of the closest-lying excited states along the Ni-N coordinate. Figure S8a shows the corresponding potential curves obtained from TDDFT calculations with the PBE0 hybrid functional. Due to the occurring triplet instabilities only the singlet manifold is considered here. Two crossings of the $\pi$-$\pi^*$ singlet state (red) with other singlet excited states – namely the $d_{z^2}$-$d^*_{x^2-y^2}$ state (green) as well as the degenerate $d_{xz}$-$d^*_{x^2-y^2}$ and $d_{yz}$-$d^*_{x^2-y^2}$ states (blue) – can be seen very close to the minimum of the $\pi$-$\pi^*$ state. The $\pi$-$\pi^*$ / $d_{xz}$-$d^*_{x^2-y^2}$ crossing is located only about 1 meV above the $\pi$-$\pi^*$ minimum, whereas the $\pi$-$\pi^*$ / $d_{z^2}$-$d^*_{x^2-y^2}$ intersection lies about 0.02 eV higher. Thus, for non-resonant excitation both crossings should be accessible, leading to a depopulation of the bright π-π* state into the dark d-d* states. We note that for the S0 ground state we found the optimized Ni-N distance at 1.918 Å (PBE0) and 1.920 Å (CAM-B3LYP), respectively. Hence, $S_0$ and $S_1$ are nested states, as usually found for phthalocyanines.

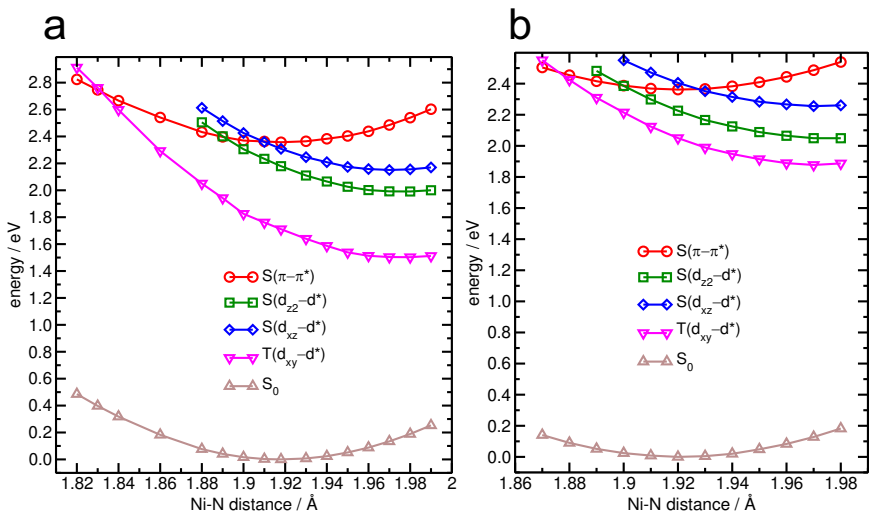

**Figure S9. TDDFT excited-state energy profiles of NiPc for varying Ni-N distances.** Here, we used the Tamm-Dancoff approximation to better assess triplet states. (a) Using the PBE0 hybrid functional; (b) using the range-separated hybrid functional CAM-B3LYP.

The exact positions of the theoretically predicted crossing points however depend on a number of factors such as the choice of exchange-correlation functional, basis set, and environment model (i.e. NaCl substrate or solvent). While we will comment on the latter below, to gauge the effect of the exchange-correlation functional, we recalculated the potential energy curves using the range-separated hybrid functional CAM-B3LYP (see Fig. S8b). We can see that the π-π* / $d_{xz}$-$d^*_{x^2-y^2}$ crossing is now slightly less accessible at 0.05 eV above the π-π* minimum, while the π-π* / $d_{z^2}$-$d^*_{x^2-y^2}$ intersection now lies only 5 meV above the π-π* minimum. From the comparison, it is already obvious that we cannot identify into which d-d* state exactly the ISC from the π-π* state occurs.

Invoking the Tamm-Dancoff approximation allows us to assess more reliably the relative energies of singlet and triplet states. Table S1 summarizes the first 17 excitations found for NiPc and which MOs are involved (see also Table S2 for orbital identification), using the PBE0 functional. Calculations as a function of Ni-N distance reveal a crossing between the π-π* singlet state and the $d_{xy}$-$d^*_{x^2-y^2}$ triplet state at 0.37 eV above the π-π* minimum (Fig. S9a). Using the CAM-B3LYP functional, on the other hand, the energy difference is only 0.11 eV (Fig. S9b). This opens up the possibility for intersystem crossing followed by radiationless relaxation back to the ground state via a number of deep-lying triplet states (Table S1) in the case of non-resonant excitation.

Concerning the effect of the environment, we assume that the flat adsorption geometry of NiPc on NaCl will likely suppress out-of-plane vibrations and also induce a friction on the Ni-N breathing mode. In addition, the interaction of NiPc with the $Na^+$ and $Cl^-$ ions will have an effect on its electronic structure and thus further change the intersection points with respect to the π-π* minimum. Besides, we note that absolute excited state energies from TDDFT generally differ from experimental values by typically 0.2-0.4 eV.[18,19] We therefore conclude that our TDDFT calculations neither permit to identify which particular d-d* state is responsible for the fast ISC from the π-π* state, nor can we quantify the ISC activation barrier in the lab setup with sufficient accuracy. Nevertheless, the TDDFT results give us a qualitative picture, rationalizing that there actually is an activation barrier for ISC due to an intersection of the π-π* singlet state with a d-d* state, as well as identifying that the "reaction coordinate" is the Ni-N bond length. We note that also the energy of the symmetric Ni-N stretch vibration at ca. 175 meV gives an indication of the minimum energy required to enable ISC from the vibrational ground state of the π-π* singlet state.

## 9. NiPc vs. HPc⁻

At first glance, there seem to be circumstantial similarities in the experimental observation of NiPc and those of deprotonated free-base phthalocyanine, i.e. HPc⁻.[20] Therefore, we would first like to comment on a potential unwanted contamination of our sample with free-base $H_2Pc$. First of all, no experiments with $H_2Pc$ have ever been performed in the UHV system used here. NiPc itself is a very stable compound, and does not dissociate by itself or in the course of sublimation (at temperatures below 350°C).[21] In order to reduce any possible $H_2Pc$ contamination of our source material, we utilized the fact that its vapor pressure is about an order of magnitude larger than that of NiPc.[22] We therefore repeatedly degassed the NiPc source material in UHV by running the Knudsen cell evaporator for up to one hour at the target deposition temperature (300-350°C). We confirmed the overall absence of $H_2Pc$ by taking a sample of the degassed material out of the crucible and performing optical spectroscopy in solution (see Section 1). From the absence of any fluorescence, we can estimate that a potential $H_2Pc$ contamination of the degassed NiPc source material would be less than 1%.

Our STM results confirm the purity of the degassed NiPc source upon cold deposition onto the NaCl/Ag(111) substrate. If there would have been $H_2Pc$ present, this would have become obvious in STM images at positive bias, as the $H_2Pc$ LUMO is not degenerate, leading to a twofold-symmetric feature in STM.[23] Furthermore, $H_2Pc$ does not deprotonate spontaneously but requires large tunneling voltages $V_s > 3.2$ V, and HPc⁻ would not be the final product, but deprotonation can continue to form $Pc^{2-}$ as well, which again has different structural, electronic and electronic and optical properties.[20] In contrast to all this, we only found one species in all our STM images, with reproducibly identical properties in STS and STML measurements, as reported here.

Finally, we would like to emphasize various differences in the observables of NiPc in our study vs. HPc⁻ reported by Vasilev *et al.*[20] The onset energies of the NiPc PIR and NIR are slightly different, leading to a gap that is 0.2 eV larger than that of HPc⁻. STM images and d$I$/d$V$ maps of the LUMO (see Fig. 1b) exhibit a fourfold symmetry, whereas HPc⁻ shows a reduced symmetry. The noise recorded in time traces of the tip-sample distance above NiPc is of a broadband nature, whereas that of HPc⁻ exhibits a four-state discrete telegraph noise due to switching of the remaing central H atom between one of the four possible isoindole $N_p$ atoms. Moreover, NiPc shows a noisy appearance in STM imaging for all voltages, whereas the HPc⁻ telegraph noise only occurred at negative voltages.

In summary, we can exclude the possibility of accidentally having measured HPc⁻, and we have confirmed the purity of our NiPc source material.

**Table S1.** TDDFT/PBE0 eigenvalues (for all excited states below 3 eV), eigenvectors, and oscillator strengths $f$ using the Tamm-Dancoff approximation for the ground-state optimized NiPc structure. Only molecular orbital contributions with a weight of more than 10% are listed. MO 141 and 142 are the HOMO and LUMO, respectively. The π-π$^*$ singlet state corresponds to #10 and #11, owing to the degenerate LUMO and LUMO+1.

| # | Spin | Energy (eV) | $f$ | MO contributions and weights |
|---|------|-------------|-----|------------------------------|
| 1 | triplet | 0.4679 | 0.0000 | 132 → 144 (98.4%) |
| 2 | triplet | 0.7796 | 0.0000 | 126 → 144 (25.4%)<br>127 → 144 (22.1%)<br>139 → 144 (24.8%)<br>140 → 144 (18.2%) |
| 3 | triplet | 0.7796 | 0.0000 | 126 → 144 (22.1%)<br>127 → 144 (25.4%)<br>139 → 144 (18.2%)<br>140 → 144 (24.8%) |
| 4 | triplet | 1.3488 | 0.0000 | 141 → 142 (87.9%)<br>141 → 143 (10.2%) |
| 5 | triplet | 1.3488 | 0.0000 | 141 → 142 (10.2%)<br>141 → 143 (87.9%) |
| 6 | triplet | 1.7110 | 0.0000 | 124 → 144 (82.8%)<br>134 → 144 (15.6%) |
| 7 | singlet | 2.1790 | 0.0000 | 132 → 144 (95.6%) |
| 8 | singlet | 2.3080 | 0.0000 | 126 → 144 (37.5%)<br>139 → 144 (47.5%) |
| 9 | singlet | 2.3080 | 0.0000 | 127 → 144 (37.5%)<br>140 → 144 (47.5%) |
| 10 | singlet | 2.3571 | 0.5797 | 141 → 142 (57.7%)<br>141 → 143 (27.1%) |
| 11 | singlet | 2.3571 | 0.5797 | 141 → 142 (27.1%)<br>141 → 143 (57.7%) |
| 12 | singlet | 2.7192 | 0.0000 | 124 → 144 (78.1%)<br>134 → 144 (20.3%) |
| 13 | triplet | 2.8144 | 0.0000 | 139 → 142 (32.7%)<br>139 → 143 (10.7%)<br>140 → 142 (10.7%)<br>140 → 143 (32.7%) |
| 14 | triplet | 2.8683 | 0.0000 | 139 → 142 (42.0%)<br>140 → 143 (42.0%) |
| 15 | triplet | 2.8921 | 0.0000 | 139 → 142 (42.4%)<br>140 → 143 (42.4%) |
| 16 | triplet | 2.9165 | 0.0000 | 141 → 144 (99.4%) |
| 17 | singlet | 2.9309 | 0.0000 | 141 → 144 (99.5%) |

**Table S2.** Molecular orbital isosurface plots (isovalue = 0.02) calculated for the optimized ground state geometry using the PBE0 functional. The selected orbitals have the highest contributions to the excited states plotted S7.

| | | | |
|---|---|---|---|
| LUMO+2 (MO 144) | 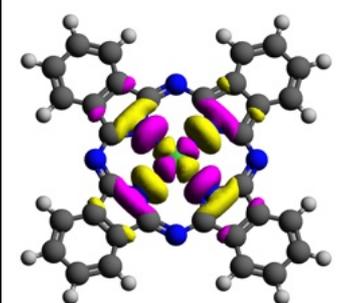 | HOMO-2 (MO 139) | 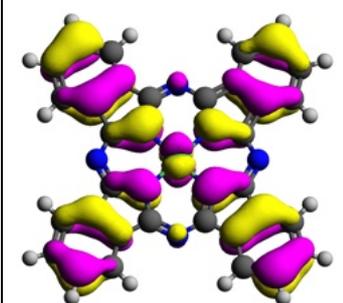 |
| LUMO+1 (MO 143) | 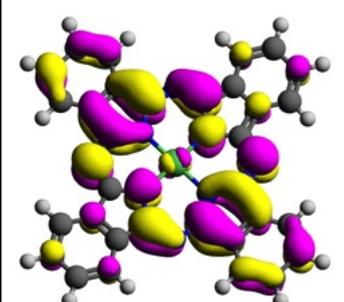 | HOMO-9 (MO 132) | 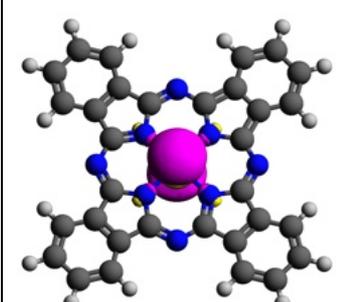 |
| LUMO (MO 142) | 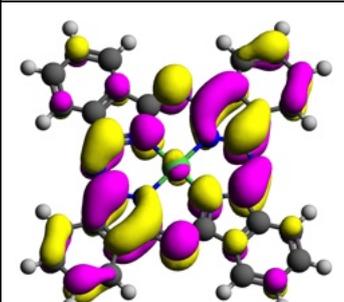 | HOMO-14 (MO 127) | 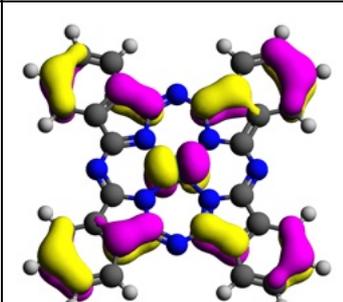 |
| HOMO (MO 141) | 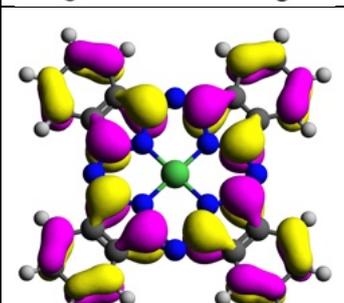 | HOMO-15 (MO 126) | 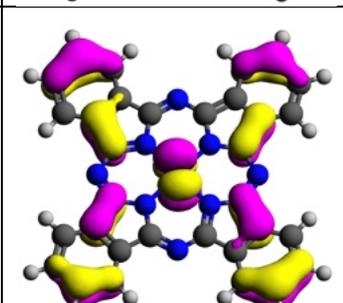 |
| HOMO-1 (MO 140) | 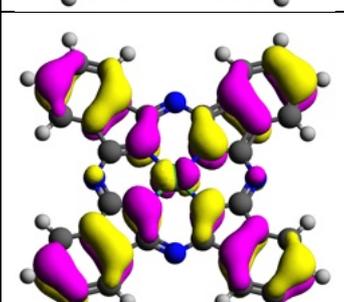 | HOMO-17 (MO 124) | 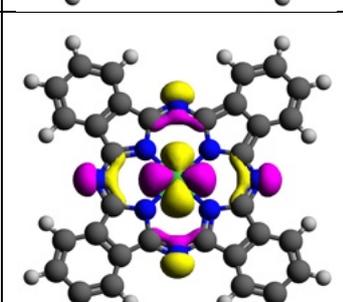 |